\documentclass[lettersize,journal]{IEEEtran}
\usepackage{amsmath,amsfonts}
\usepackage{algorithmic}
\usepackage{algorithm}
\usepackage{array}
\usepackage[caption=false,font=normalsize,labelfont=sf,textfont=sf]{subfig}
\usepackage{textcomp}
\usepackage{stfloats}
\usepackage{url}
\usepackage{verbatim}
\usepackage{graphicx}
\usepackage{cite}
\usepackage{booktabs}
\usepackage{multirow}
\usepackage{hyperref}
\usepackage{caption}
\hypersetup{
	colorlinks=true,
	linkcolor=blue,
	citecolor=blue,
}

\usepackage{amsmath,amsfonts,bm}

\def\eqref#1{equation~\ref{#1}}

\def\1{\bm{1}}

\DeclareMathAlphabet{\mathsfit}{\encodingdefault}{\sfdefault}{m}{sl}
\SetMathAlphabet{\mathsfit}{bold}{\encodingdefault}{\sfdefault}{bx}{n}

\begin{document}

\title{ScentGen: Hierarchical Multimodal Olfactory Semantic Modeling for Molecular Odor Description Generation}

\author{
Zhiliang Wu$^{*}$, \emph{Member, IEEE},
Zhaolin Hu$^{*}$,
Hehe Fan, \emph{Senior Member, IEEE},
and Weisi Lin, \emph{Fellow, IEEE}
\thanks{
Manuscript received 30 July 2026. This project was supported by the Agency for Science, Technology, and Research (A*STAR) under its MTC Programmatic Funding Scheme (M23L8b0049) Scent Digitalization and Computation Program, and the Earth System Big Data Platform of the School of Earth Sciences, Zhejiang University. \emph{(Corresponding author: Weisi Lin)}}
\IEEEcompsocitemizethanks
{
\IEEEcompsocthanksitem 
Zhiliang Wu and Weisi Lin are with the College of Computing and Data Science, Nanyang Technological University, Singapore 639798, Singapore (e-mail: zhiliang.wu@ntu.edu.sg and wslin@ntu.edu.sg).
\IEEEcompsocthanksitem
Zhaolin Hu and Hehe Fan are with the School of Artificial Intelligence, Zhejiang University, Hangzhou 310007, China (e-mail: 12321165@zju.edu.cn and hehefan@zju.edu.cn).\\
\textsuperscript{*}These authors contributed equally to this work.
}
}
\markboth{IEEE Transactions on Multimedia,~Vol.~XX, No.~XX, Month~2026}%
{YourLastName \MakeLowercase{\textit{et al.}}: Short Title of Your Paper}


\maketitle

\begin{abstract}
In this paper, we introduce a molecular odor description generation task, which aims to generate natural language odor descriptions from molecular structures.
Unlike conventional methods that describe molecular odor using discrete labels, this task generates expressive and human-interpretable sensory descriptions. 
To address this task, we propose a hierarchical multimodal olfactory semantic modeling framework, named ScentGen. 
ScentGen consists of three key components: an odor semantic planner, a semantic adapter, and a description generator. 
The odor semantic planner integrates complementary molecular information from 1D SMILES sequences, 2D molecular graphs, and 3D molecular conformations to learn discriminative and structured olfactory semantics. 
The semantic adapter further maps the learned olfactory representation into the hidden space of a large language model, transforming molecular odor semantics into language-compatible continuous prompts.
Conditioned on these prompts, the description generator produces coherent odor descriptions that reflect plausible sensory characteristics of the input molecule.
Considering the lack of molecular datasets with natural language odor descriptions, we further construct a molecular odor description dataset containing paired multimodal molecular representations and human-interpretable odor descriptions. 
Extensive experiments demonstrate that ScentGen generates coherent and expressive odor descriptions, providing a more flexible solution for molecular odor understanding beyond discrete odor label prediction. 
Project Page: \href{https://wzlwlm13.github.io/ScentGen/}{https://scentgen.github.io}.
\end{abstract}

\section{Introduction}
\IEEEPARstart{O}{dor} is a semantically rich yet computationally challenging sensory modality~\cite{lee2023principal,xie2025multi,4797804,su2026nose}.
Unlike images or sounds, which can be directly captured and represented as pixels or waveforms, odor cannot be characterized by a single physical signal~\cite{aktas2024odor,11194260,mcconachie2025low}. 
Its perception arises from the complex interplay among molecular structure, physicochemical properties, olfactory receptors, and subjective human experience.
A molecule may evoke multiple perceptual attributes, such as {fruity}, {sulfurous}, and {green}, while structurally similar molecules can produce markedly different smells~\cite{zhang2024deep}. 
These characteristics make molecular odor understanding a challenging cross-modal problem, where chemical structures need to be mapped to human-interpretable sensory descriptions.

\begin{figure}[!t]
    \centering
    \includegraphics[width=1\linewidth]{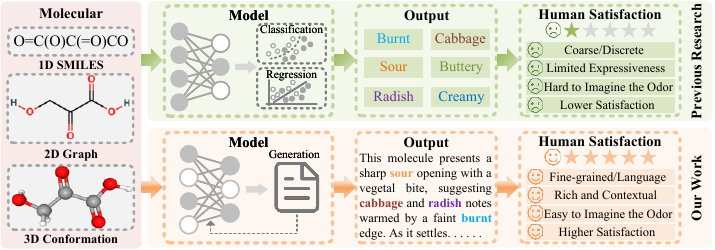}
    \vspace{-12pt}
    \caption{
    {Comparison between conventional molecular odor label prediction and our molecular odor description generation task. 
    Our task moves beyond coarse discrete labels and generates fine-grained natural language descriptions for more expressive and human-understandable odor perception.}}
    \label{fig_i}
    \vspace{-12pt}
\end{figure}

Recent advances in artificial intelligence (AI)~\cite{11285488} have greatly advanced molecular odor modeling~\cite{bierling2025dataset}, particularly in predicting odor-related labels from molecular structures. 
Early studies~\cite{keller2017predicting,li2018accurate,suh2025comparative}, such as the DREAM olfaction prediction challenge~\cite{keller2017predicting}, demonstrated that machine learning models can infer human olfactory perceptual attributes from chemical features, including intensity, pleasantness, and semantic descriptors.
Subsequently, deep learning methods~\cite{zhang2025atomas,iwata2025interpretable,zhang2024deep,taleb2024can}, especially graph neural networks (GNNs), were introduced to directly model structure–odor relationships from molecular graphs, thereby improving odor prediction accuracy.
More recently, large-scale olfactory datasets and odor representation spaces have further supported data-driven olfaction modeling.
For example,
Pyrfume~\cite{hamel2024pyrfume} provides large-scale olfactory data, while the principal odor map (POM)~\cite{lee2023principal} offers a structured representation space for modeling odor perception.
These studies show that AI models can capture meaningful associations between molecular structures and olfactory semantics.


However, existing methods~\cite{saini2022predicting,keller2017predicting,su2026nose,lee2023principal} usually formulate molecular odor prediction as a classification or regression problem over a predefined set of discrete odor labels.
While such a paradigm is useful for benchmarking and high-throughput screening, it does not fully reflect how odors are described in real-world applications~\cite{bierling2025dataset}, as shown in Fig.~\ref{fig_i}.
In perfumery, flavor chemistry, food science, and sensory evaluation, odors are often expressed through natural language descriptions, such as “a fruity and sweet aroma with a slightly green undertone”.
Compared with discrete labels, natural language descriptions can convey richer sensory information, including subtle nuances, relative prominence, and perceptual relations~\cite{zhong2024bridging}.
For example, annotating a molecule with both “fruity” and “sulfurous” only provides a coarse indication of its odor attributes.
It remains unclear whether the sulfurous note dominates or is subtle, whether the fruity impression resembles pineapple, apple, or another fruit, and how these notes interact to form a coherent sensory profile.



Meanwhile, large language models (LLMs)~\cite{li2024empowering,YOU2025174,su2026nose,edwards2021text2mol,edwards2022translation,zhao2025developing} provide new opportunities for generating fluent and domain-specific scientific descriptions. 
In chemistry, recent studies~\cite{edwards2021text2mol,edwards2022translation,zhao2025developing} have begun to explore molecule–language modeling, where molecular structures are represented, retrieved, or generated together with natural language descriptions. 
For instance, Text2Mol~\cite{edwards2021text2mol} explored cross-modal retrieval between molecular structures and textual descriptions, demonstrating that the two modalities can be aligned in a shared semantic space. 
MolT5~\cite{edwards2022translation} further extended this paradigm to generative tasks, including molecule captioning and text-guided molecular generation. 
While these studies collectively establish the feasibility of aligning molecular structures with natural language, they mainly focus on general molecule–language modeling that describes chemical structures or physicochemical properties, leaving the perceptual dimension of odor underexplored.
In contrast, odor description generation poses a greater challenge, as it requires translating molecular cues into subjective human sensory impressions.
The generated descriptions should be grounded in odor-aware semantics, capturing plausible olfactory attributes, subtle variations in intensity, and the sensory relations among individual notes.


In this paper, we introduce a  molecular odor description generation task, which aims to generate natural language odor descriptions directly from molecular structures. 
Unlike conventional molecular odor label prediction, this task requires the model to produce expressive and human-interpretable sensory descriptions. 
Such descriptions need to capture not only individual odor attributes, but also their combinations, relative prominence, and semantic relations.
To this end, we propose a hierarchical multimodal olfactory semantic modeling framework, named ScentGen.
ScentGen consists of three key components: an odor semantic planner, a semantic adapter, and a description generator. 
The odor semantic planner first learns discriminative and structured olfactory semantics from multimodal molecular inputs.
By integrating complementary information from the 1D SMILES sequence, the 2D molecular graph, and the 3D molecular conformation, it captures odor-relevant cues from different perspectives and produces a unified olfactory representation. 
Then, the semantic adapter maps the unified olfactory representation into the hidden space of an LLM, transforming molecular odor semantics into language-compatible continuous prompts. 
Finally, the description generator produces coherent natural language descriptions from these prompts, thereby reflecting the plausible odor characteristics of the input molecule.
Through this hierarchical design, ScentGen bridges multimodal molecular representations and natural language odor descriptions, enabling more accurate and expressive molecular odor description generation.

Considering the lack of molecular datasets with natural language odor descriptions, we further construct a molecular odor description dataset to support this task. 
The dataset provides paired molecular structures and human-interpretable odor descriptions, enabling systematic training and evaluation of molecular odor description generation models. 
Extensive experiments demonstrate that ScentGen moves beyond fixed odor label prediction and provides a more expressive solution for molecular odor understanding.


Our main contributions can be summarized as follows:
\begin{itemize}
\item We formulate molecular odor description generation, which produces natural language odor descriptions directly from molecular structures. 
This task extends molecular odor modeling beyond predefined discrete-label prediction toward expressive, human-interpretable sensory description generation.

 \item We propose ScentGen, a hierarchical multimodal olfactory semantic modeling framework. 
 It uses an odor semantic planner to integrate complementary multimodal molecular information, and further employs a semantic adapter to convert the learned olfactory semantics into language-compatible prompts, thereby guiding the description generator to produce coherent odor descriptions.

 \item We construct a molecular odor description dataset. 
 The dataset provides paired multimodal molecular representations and natural language odor descriptions.
 The dataset will be released to facilitate future research on molecular odor description generation.
\end{itemize}
\section{Related Work}
\subsection{Artificial Intelligence for Molecular Odor Prediction}
In the field of single-molecule olfactory perception modeling, early studies mainly followed the quantitative structure-odor relationship paradigm. These methods~\cite{saini2022predicting,snitz2013predicting,keller2017predicting,licon2019chemical,suh2025comparative} typically extract physicochemical descriptors, structural fragments, or molecular fingerprints, and then feed them into conventional machine learning models to predict odor categories.
However, their reliance on hand-crafted features limits their ability to capture label correlations and the complex nonlinear mapping between molecular structure and odor.

Driven by the rapid progress of deep learning, recent studies have moved beyond hand-crafted descriptors toward data-driven representation learning for molecular odor prediction.
Keller et al.~\cite{keller2017predicting} were among the first to systematically demonstrate that human perception of single-molecule odors can be predicted from chemical features, thereby establishing a standardized data-driven setting for this task. 
Building on this foundation,
Sharma et al.~\cite{sharma2021smiles} used DNNs and CNNs to automatically learn odor-related representations from physicochemical properties, molecular fingerprints, and structural images. 
Saini et al.~\cite{saini2022predicting} further formulated molecular odor prediction as a multi-label learning problem, emphasizing the importance of modeling correlations among co-occurring odor descriptors.
Beyond category-level prediction, subsequent work has shifted toward more generalizable representation learning. 
A representative work is the POM~\cite{lee2023principal}, which leverages GNNs to construct a unified olfactory space.
This representation generalizes better to unseen molecules and transfers more effectively across diverse perceptual tasks.
More recent studies have incorporated richer molecular cues, such as 3D conformations~\cite{iwata2025interpretable}, electrostatic information~\cite{zhang2024deep}, and pretrained molecular Transformers~\cite{taleb2024can}.

Despite these advances, they still rely on predefined odor labels to characterize molecular smells.  
Such fixed labels cannot fully express fine-grained perceptual differences, mixed odor impressions, or the rich semantics of olfactory experience. 
This motivates the development of methods that generate natural language odor descriptions, enabling more comprehensive and human-aligned molecular olfactory modeling.

\subsection{Molecule-to-Text Generation}
Recent studies on molecule-to-text generation have gradually shifted from molecule-text alignment toward generative molecular semantic modeling. 
Early works such as Text2Mol~\cite{edwards2021text2mol} and MolT5~\cite{edwards2022translation} established the basic paradigm of molecule captioning from the perspectives of cross-modal retrieval and sequence-to-sequence pretraining. 
Subsequent methods further enriched this paradigm from different perspectives. 
For example, MolReGPT~\cite{li2024empowering} introduced retrieval-augmented in-context learning, 3D-MoLM~\cite{li2024towards} and 3D-MolT5~\cite{pei20243d} incorporated 3D structural information, and Atomas~\cite{zhang2025atomas} explored fine-grained fragment-text alignment.
Although these studies have improved the structural awareness and semantic expressiveness of molecule-to-text generation, they remain focused on the intrinsic chemical semantics of a molecule (\emph{e.g.}, its structure, composition, and physicochemical or functional properties), failing to generate fine-grained and human-like odor descriptions.
Therefore, generating rich, perception-based odor descriptions remains an open problem.

\begin{table}[tb]
  \centering
  \footnotesize
  \renewcommand\tabcolsep{3pt}
  \caption{Comparison of molecule-odor prediction datasets.} 
  \vspace{-0.1cm}
    \begin{tabular}{c||cccccccc}
    \hline
    \hline
    \textbf{Dataset} & \textbf{SMILES} & \textbf{Graph} & \textbf{Conformation} & \textbf{Text} & \textbf{Number} \\
    \hline
    \hline
Leffingwell~\cite{sanchez2019machine}        & $\checkmark$ & $\times$     & $\times$   & $\times$     & 3,561 \\
FlavorNet~\cite{arn1998flavornet}          & $\checkmark$ & $\times$     & $\times$     & $\times$     & 718 \\
Keller~\cite{keller2016olfactory}         & $\checkmark$ & $\times$     & $\times$     & $\times$     & 480 \\
DREAM~\cite{li2018accurate}              & $\checkmark$ & $\times$     & $\times$    & $\times$     & 476 \\
OlfactionBase~\cite{sharma2022olfactionbase}      & $\checkmark$ & $\times$     & $\times$     & $\times$     & 5,109 \\
Subset-IGD~\cite{ameta2025odor}         & $\checkmark$ & $\times$     & $\times$   & $\times$     & 2,606 \\
OpenPOM~\cite{lee2023principal}            & $\checkmark$ & $\times$ & $\times$      & $\times$     & 4,983 \\
Ours               & $\checkmark$ & $\checkmark$ & $\checkmark$ & $\checkmark$ & 4,983 \\
    \hline
    \hline
    \end{tabular}
  \label{EA}
  \vspace{-0.2cm}
\end{table}
\section{Method}
\subsection{Problem Formulation}
Given a molecule, our goal is to generate a natural language description that is linguistically fluent and faithfully reflects the molecule's odor characteristics.
Existing molecular odor prediction methods mainly focus on predicting a set of discrete odor descriptors, \emph{e.g.}, fruity, floral, or woody.
While such descriptors offer useful semantic cues, they characterize only isolated olfactory attributes and cannot convey a molecule's overall sensory profile in natural language. 
In contrast, we study molecule-grounded odor description generation, where the model is required to generate a complete odor description conditioned on multimodal molecular information.

Formally, given a molecule $\mathcal{M}=\{S,G,C\}$, 
$S$ is a one-dimensional (1D) SMILES sequence that encodes atom-level sequential information.
$G$ is a two-dimensional (2D) molecular graph that captures the topological connectivity among atoms.
$C$ is a three-dimensional (3D) molecular conformation that provides spatial geometric information about molecular shape and interactions. 
The goal of molecular odor description is to generate a complete odor description $\widehat{Y}=\{\widehat{y}_1,\widehat{y}_2,\dots,\widehat{y}_m\}$, 
which is fluent, structurally grounded, and semantically aligned with the ground-truth description $Y=\{y_1,y_2,\ldots,y_n\}$. 
$\widehat{y}_i$ and ${y}_i$ denote the $i$-th token of the generated and ground-truth descriptions, and $m$ and $n$ denote their corresponding lengths.

\begin{figure}[!t]
    \centering
    \includegraphics[width=1\linewidth]{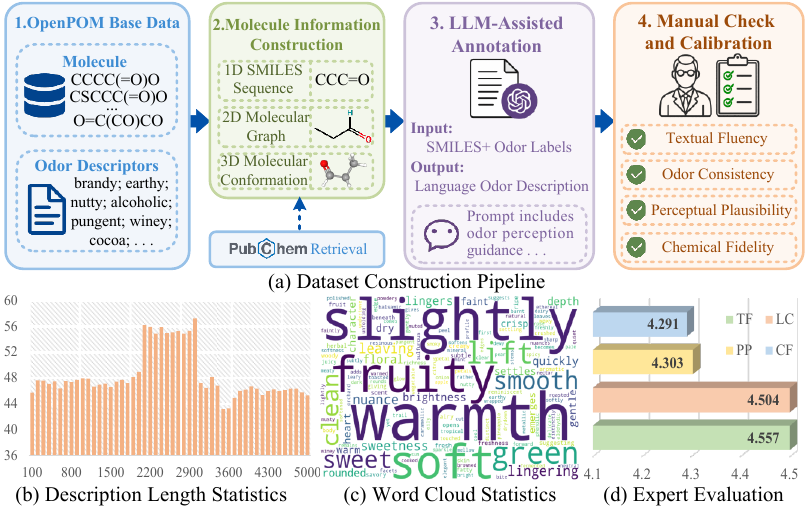}
    \vspace{-15pt}
    \caption{
    {Dataset construction pipeline.}
    }
    \label{fig_data}
    \vspace{-10pt}
\end{figure}

\subsection{Molecular Odor Description Dataset}

Existing molecular odor datasets are mainly designed for discrete odor label prediction, such as fruity, floral, or woody~\cite{keller2017predicting,saini2022predicting,li2018accurate,zhang2024deep}. 
Although these datasets have advanced molecular odor prediction, they are not well suited to molecular odor description generation task. 
As shown in Table~\ref{EA}, on the one hand, they usually lack complete natural language odor descriptions. 
This makes it hard to train and evaluate models for fluent and coherent odor description generation. 
On the other hand, they often provide limited molecular information. 
However, odor perception is affected by multiple molecular cues, including 1D SMILES sequences, 2D molecular topology, and 3D spatial conformations~\cite{hamel2024pyrfume}.

To support this task, we construct a molecular odor description dataset based on OpenPOM~\cite{lee2023principal}, termed MOD.
The MOD dataset contains richer molecular information and natural language odor descriptions.
As shown in Fig.~\ref{fig_data}(a), we first collect molecules and odor labels from OpenPOM.
For each molecule, we retain its SMILES sequence and original odor labels, and then retrieve its 2D molecular graph and 3D molecular conformation from PubChem~\cite{kim2025pubchem} using the SMILES sequence.
In this way, each molecule is represented by three complementary modalities, including the 1D SMILES sequence, the 2D molecular graph, and the 3D molecular conformation. 
These modalities provide comprehensive molecular cues for odor description generation.
Based on these molecular cues, we further adopt an LLM-assisted (GPT) annotation strategy to generate natural language odor descriptions.
Unlike existing datasets that provide only discrete odor labels, our descriptions characterize odor perception in a continuous semantic form, covering both the dominant odor profile and its secondary and subtle nuances.
To ensure annotation quality, all generated descriptions are manually checked and revised by chemistry experts.
This process keeps the descriptions consistent with the original odor labels, improves linguistic fluency, and enhances chemical plausibility while correcting potential errors.
As a result, the constructed MOD dataset provides both multimodal molecular information and high-quality odor descriptions, supporting a more fine-grained and realistic representation of molecular odor characteristics.

The constructed MOD dataset contains $4,983$ molecular odor descriptions, including $4,484$ for training and $499$ for testing. 
As shown in Fig.~\ref{fig_data}(b), the description length generally ranges between $40$ and $60$ words. 
This indicates that the descriptions are neither so short that they miss important odor cues nor so long that they introduce redundant information.
Fig.~\ref{fig_data}(c) further presents the word cloud statistics of the descriptions. 
High-frequency words such as fresh, fruity, warm, soft, green, sweet, and smooth show that the dataset covers diverse and meaningful odor semantics.
These statistics demonstrate that the MOD dataset contains rich sensory expressions and can support the learning of fine-grained odor description patterns.
To further validate the quality of the descriptions, we conduct an expert evaluation.
We design a questionnaire and present the SMILES sequence and a set of discrete odor labels of each molecule. 
Experts then rate the corresponding odor description on four dimensions: \emph{textual fluency}, \emph{label coverage}, \emph{perceptual plausibility}, and \emph{chemical fidelity}. 
We randomly sample the descriptions of $50$ molecules and invite $15$ experts with backgrounds  in chemistry or biology.
Each expert scores every dimension on a five-point scale, from $1$ (lowest quality) to $5$ (highest quality).
As shown in Fig.~\ref{fig_data}(d), the descriptions receive consistently high ratings across all four dimensions.
The average scores are $4.557$ for textual fluency, $4.504$ for label coverage, $4.303$ for perceptual plausibility, and $4.291$ for chemical fidelity.
These results show that the odor descriptions in the MOD dataset are fluent in language, faithful to the given odor labels, perceptually plausible, and chemically consistent with the molecules. 
They further confirm the reliability of the dataset.

\begin{figure}[!t]
    \centering
    \includegraphics[width=0.98\linewidth]{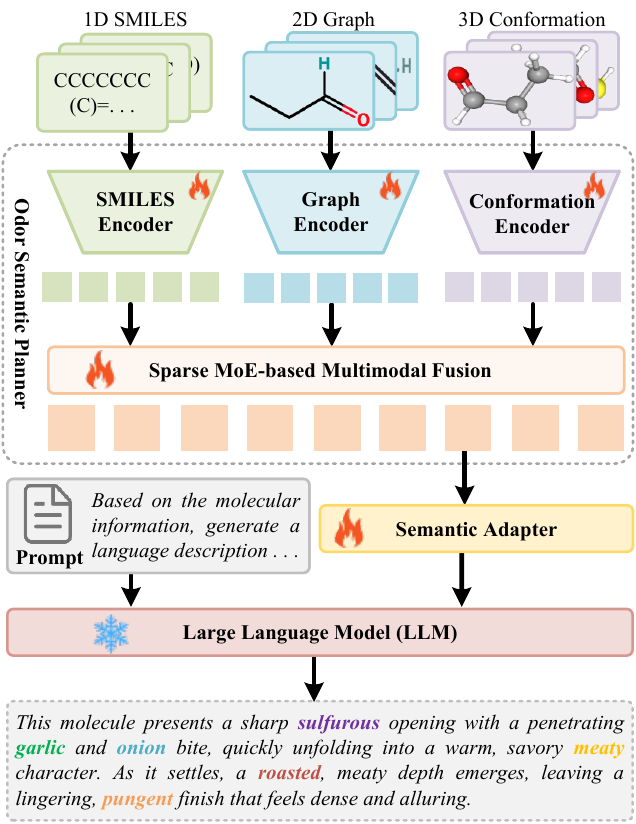}
    \caption{
    {\textbf{Overview of the proposed ScentGen framework}. Given the 1D SMILES, 2D molecular graphs, and 3D conformations of a molecule, ScentGen first extracts compact olfactory semantics through an odor semantic planner. A semantic adapter then transforms the planned semantics into continuous soft prompts, guiding the LLM-based odor description generator to produce fluent, coherent, and sensory-faithful odor descriptions.
    }}
    \label{fig_kt}
    \vspace{-12pt}
\end{figure}

\subsection{Odor Description Generation Framework}

As shown in Fig.~\ref{fig_kt}, we propose a hierarchical multimodal olfactory semantic modeling framework for molecular odor description generation, named ScentGen. 
ScentGen consists of three core components: an odor semantic planner, a semantic adapter, and an LLM-based description generator.
The odor semantic planner first extracts odor-related cues from the molecular representation and organizes them into compact olfactory semantics, thereby establishing a semantic bridge between molecular structure and odor language. 
This design makes the generation process more controllable and odor-aware. 
Next, the semantic adapter converts these olfactory semantics into continuous soft prompts that the LLMs can directly process. 
Compared with discrete odor labels, such soft prompts retain richer semantic relations and mitigate the risk of error propagation caused by hard-label guidance.
Finally, the LLM-based description generator takes both the adapted semantic prompts and the textual instruction as input, and generates fluent and coherent odor descriptions consistent with the molecular structure.
Through this hierarchical design, ScentGen progressively converts molecular information into olfactory semantics and then into natural language, enabling more reliable and sensory-faithful odor description generation.

\subsubsection{\textbf{Odor Semantic Planner}}
The odor semantic planner serves as a semantic bridge between molecular structures and odor description generation. 
It extracts compact and informative olfactory semantics from molecules, providing structured guidance for the downstream description generator. 
As shown in Fig.~\ref{fig_kt}, the planner consists of three modality-specific encoders and a sparse Mixture-of-Experts (MoE) feature fusion module.
The three encoders extract odor-related cues from 1D SMILES sequences, 2D molecular graphs, and 3D conformations, respectively, while the fusion module adaptively integrates these cues by retaining useful information and reducing redundancy.

{\textbf{1D SMILES encoder}} 
is designed to capture odor-related sequential cues from molecular string representations.
Compared with molecular graphs and conformations, a SMILES sequence provides a compact description of chemical syntax, which helps model atom order, branches, rings, and functional patterns in a linear form. 
We adopt a character-level SMILES Transformer~\cite{honda2019smiles,mswahili2024transformer} as the encoder to learn sequence-aware molecular features from SMILES  sequences.
In this way, the encoder captures both local chemical patterns and long-range dependencies among chemical symbols, providing complementary sequential cues for odor description generation.

{\textbf{2D graph encoder}}
aims to capture odor-related topological cues from molecular graphs. 
Compared with the 1D SMILES sequence, a molecular graph explicitly describes atom–bond connectivity, which helps model functional groups and local chemical substructures.
We adopt a four-layer GINE network~\cite{hu2020strategies} as the graph encoder. 
Unlike conventional GNNs~\cite{hu2026oscagent,hu2026graphtarif} that mainly aggregate neighboring node features, GINE incorporates edge features into message passing, enabling the model to better capture bond-dependent atom interactions. 
In this way, the encoder learns topology-aware molecular representations that complement the sequential cues from the SMILES branch.

{\textbf{3D conformation encoder}} 
is used to capture geometry-aware odor cues from molecular conformations.
We employ a four-layer EGNN~\cite{satorras2021n,jian2025reaction} to model local spatial interactions among atoms. 
By incorporating relative interatomic distances into message passing, the encoder characterizes the spatial arrangement of atoms and functional groups. 
Its equivariant design further keeps the learned representations consistent under rotations and translations, which is important for modeling molecular geometry~\cite{jiao2026equivariant,huang2024learning}.
In this way, the encoder learns robust geometric representations and provides complementary spatial cues for odor description generation.

{\textbf{Sparse MoE-based feature fusion module}} 
aims to adaptively integrate complementary cues from different molecular modalities. 
After obtaining the embeddings from the 1D SMILES, 2D molecular graph, and 3D conformation encoders, we project them into a shared $256$-dimensional semantic space to ensure consistent feature dimensions. 
The projected features are then fed into a sparse MoE-based fusion module. 
In this module, a softmax-based router predicts the importance weights of each expert for a given molecule. 
Instead of activating all experts, the router selects only the top two experts with the highest routing weights~\cite{cai2025survey,lin2026moe}. 
The outputs of the selected experts are weighted by their corresponding routing scores and summed to generate the final fused olfactory semantic features. 
In this way, the model learns molecule-specific fusion patterns, dynamically emphasizing more informative modality cues while reducing the influence of redundant or less relevant information. 
Such a design also avoids unnecessary computation, making the fusion process efficient and flexible.

\subsubsection{\textbf{Semantic Adapter}}
The semantic adapter is designed to bridge the olfactory semantic representation produced by the odor semantic planner and the embedding space of the frozen LLM. 
The odor semantic planner integrates information from 1D SMILES sequences, 2D molecular graphs, and 3D conformations, yielding a compact $256$-dimensional MoE-fused feature that encodes molecule-specific odor semantics. 
However, this representation remains a continuous molecular feature, whereas the LLM operates on discrete token embeddings.
Therefore, it cannot be directly used as a textual prompt.

To bridge this gap, we introduce a semantic adapter. 
Such an adapter projects the fused olfactory representation into a higher-dimensional prompt space and reshapes it into a sequence of continuous virtual prompt tokens.
These tokens are then inserted into the LLM embedding space as a molecule-aware semantic prefix. 
In this way, the adapter converts compact molecular odor semantics into token-like continuous prompts, providing a flexible and structured interface for odor description generation.
This design offers two main advantages. 
First, the adapter establishes a direct connection between molecular representation learning and language generation. 
Instead of converting molecular semantics into handcrafted natural language prompts, it learns to represent odor-related information in the continuous embedding space of the LLM. 
This helps retain fine-grained olfactory cues from different molecular modalities, including subtle sensory information that a small set of fixed odor words may fail to express.
Second, the LLM remains frozen during this process, so its general language ability can be preserved. 
The semantic adapter only learns to translate molecular odor semantics into LLM-compatible prompt embeddings, making the conditioning process both stable and efficient.
During generation, the LLM receives both the molecule-aware semantic prefix and the textual instruction, enabling it to produce fluent, coherent odor descriptions consistent with the input molecule.

\subsubsection{\textbf{Description Generator}}
The description generator aims to generate natural language odor descriptions from molecule-aware semantic prompts. 
In this paper, we adopt Qwen-8B as the generator. 
After the semantic adapter transforms the fused olfactory representation into continuous virtual prompt tokens, these tokens are concatenated with a task instruction and inserted into the embedding space of Qwen-8B. 
This allows the generator to receive molecule-specific odor semantics in a form compatible with language generation. 
Conditioned on the semantic prompts and the task instruction, Qwen-8B produces fluent, coherent, and sensory-consistent odor descriptions for the input molecule. 
Compared with simple odor label prediction, this paradigm enables the model to generate more expressive and fine-grained descriptions of molecular odors.

\subsection{Loss Function}
The overall loss function of ScentGen consists of two parts: the semantic planning loss $\mathcal{{L}}_{Pla}$ and the odor description generation loss $\mathcal{{L}}_{Gen}$. 
The former guides the odor semantic planner to learn discriminative and structured olfactory semantics from molecular representations, while the latter optimizes the semantic adapter to convert the learned olfactory semantics into language-compatible continuous prompts for the LLM.
The overall loss can be expressed as: 
\begin{equation}
\mathcal{{L}}=\mathcal{{L}}_{Pla}+\mathcal{{L}}_{Gen}.
\label{pla1}
\end{equation}

\subsubsection{\textbf{Semantic Planning Loss $\mathcal{{L}}_{Pla}$}}
To better constrain the semantic planner to learn discriminative and structured olfactory semantics from different molecular modalities, we introduce a classification head on top of its output for discrete odor label prediction.  
This auxiliary prediction task provides explicit odor-level supervision for the planner, encouraging the learned representation to capture label-discriminative olfactory cues. 
Considering that odor labels are usually sparse and highly imbalanced, we first adopt the Asymmetric Loss (ASL) $\mathcal{{L}}_{ASL}$~\cite{ridnik2021asymmetric} to supervise multi-label odor prediction, which assigns different focusing mechanisms to positive and negative samples and has been shown to be effective for highly imbalanced multi-label classification tasks~\cite{chen2025towards,huang2025dual}. 
To further model the co-occurrence structure among odor labels, we introduce a batch-level cosine label correlation loss $\mathcal{{L}}_{Cor}$. 
By minimizing the discrepancy between the predicted label relevance and the ground-truth label relevance, $\mathcal{L}_{Cor}$ encourages the predicted label correlations to match the ground-truth label relationships.
$\mathcal{L}_{Cor}$ is defined as follows:
\begin{equation}
\mathcal{L}_{Cor}=\frac{1}{M}\sum_{i\neq{j}}\Big(\mathcal{C}(\widehat{p}_i,\widehat{p}_j)-\mathcal{C}({p}_i,{p}_j)\Big)^2,
\label{fusion12}
\end{equation}
where $\widehat{p}_i$, $\widehat{p}_j$, ${p}_i$, and ${p}_j$ denote the $i$-th and $j$-th predicted odor label and ground-truth label, respectively.
$\mathcal{C}(\cdot,\cdot)$ is the correlation between labels calculated by the inner product between label vectors. $M$ is the number of off-diagonal elements in the correlation matrix.
Furthermore, we apply a load-balancing regularization term $\mathcal{{L}}_{MoE}$~\cite{fedus2022switch,lin2026moe} to the MoE routing mechanism to prevent expert collapse and encourage more balanced expert utilization.
In this design, the semantic planning loss $\mathcal{{L}}_{Pla}$ can be formulated as:
\begin{equation}
\mathcal{{L}}_{Pla}=\mathcal{{L}}_{ASL}+\lambda_{Cor}\mathcal{{L}}_{Cor}+\lambda_{MoE}\mathcal{{L}}_{MoE},
\label{pla2}
\end{equation}
where $\lambda_{Cor}$ and $\lambda_{MoE}$ are non-negative trade-off parameters. 

\begin{table*}[t]
\centering
\footnotesize
\caption{Quantitative comparison on molecular odor description generation.}
\label{tab_quantitative_results}
\vspace{-5pt}
\begin{tabular}{lcccccc}
\hline
\hline
\multirow{2}{*}{Method} 
& \multicolumn{4}{c}{Text Semantic Level} 
& \multicolumn{2}{c}{Odor Label Level} \\
\cmidrule(lr){2-5} \cmidrule(lr){6-7}
& BLEU$\uparrow$ & METEOR$\uparrow$ & ROUGE$\uparrow$ & BERTScore-F1$\uparrow$ 
& FCM$\uparrow$ & F1$\uparrow$ \\
\hline
\hline
GLM-5.1        & 5.710  & 0.204 & 0.234 & 0.738 & 0.527 & 0.535 \\
Qwen-8B        & 0.189 & 0.069 & 0.126 & 0.695 & 0.432 & 0.477 \\
Qwen-3.6        & 4.520  & 0.181 & 0.242 & 0.745 & 0.558 & 0.558 \\
DeepSeek-V4    & 3.730  & 0.175 & 0.247 & 0.751 & 0.549 & 0.564 \\
Gemini-3.1      & 1.480  & 0.184 & 0.220 & 0.744 & 0.538 & 0.549 \\
Claude-Opus  & 2.390  & 0.202 & 0.240 & 0.747 & 0.587 & 0.571 \\
GPT-4.1        & 0.991 & 0.157 & 0.210 & 0.734 & 0.509 & 0.522 \\
GPT-5.5        & 5.416 & 0.201 & 0.281 & 0.755 & 0.572 & 0.575 \\
ChemDFM        & 0.073 & 0.077 & 0.127 & 0.636 & 0.099 & 0.113 \\
SFT            & 14.200  & 0.330 & 0.320 & 0.768 & 0.549 & 0.542 \\
Ours           & \textbf{15.500} & \textbf{0.351} & \textbf{0.353} & \textbf{0.776} & \textbf{0.605} & \textbf{0.616} \\
\hline
\hline
\end{tabular}
\vspace{-8pt}
\end{table*}

\subsubsection{\textbf{Description Generation Loss $\mathcal{{L}}_{Gen}$}}
To align the continuous olfactory prompts with natural language odor descriptions, we introduce a description generation loss $\mathcal{{L}}_{Gen}$. 
After the semantic planner produces the fused olfactory representation, the semantic adapter maps it into a sequence of virtual prompt tokens $h_{mol}$. 
These tokens are concatenated with the task instruction $z_{prompt}$ and then fed into the frozen LLM to generate the target odor description.
In this scenario, we constrain the generated description with $\mathcal{{L}}_{Gen}$, encouraging the semantic adapter to produce language-compatible olfactory cues that can be better understood and utilized by the frozen LLM.
The loss $\mathcal{{L}}_{Gen}$ is defined as follows:
\begin{equation}
\mathcal{{L}}_{Gen}=-\frac{1}{n}\sum_{t=1}^nlog~p({y}_t|h_{mol},z_{prompt},{y}_{<t}),
\label{pla}
\end{equation}
where $y_t$ is $t$-th token in the ground-truth description.
$n$ denotes the number of tokens in the description.

\section{Experiments}
\subsection{Baselines and Metrics}
We compare our method with three groups of baselines.
The first group comprises general LLMs, including GLM-$5.1$, Qwen-$8$B, Qwen-$3.6$, DeepSeek-V$4$, Gemini-$3.1$, Claude-Opus, GPT-$4.1$, and GPT-$5.5$. 
These models are evaluated under the same setting for generating molecular odor descriptions to assess the capability of general LLMs in producing odor-related descriptions. 
The second group consists of the chemistry-oriented language model ChemDFM~\cite{zhao2025developing}, which tests the benefit of chemical-domain knowledge for this task.
The third group is a supervised fine-tuning (SFT) baseline, where Qwen-$8$B is directly fine-tuned on our constructed MOD dataset.
For fair comparison, all baselines receive identical prompts and generate natural-language odor descriptions.

We evaluate the generated descriptions at two complementary levels: \emph{text-semantic level} and \emph{odor-label level}.
At the text-semantic level, we adopt BLEU~\cite{li2024towards}, METEOR~\cite{pei20243d}, ROUGE~\cite{li2024towards}, and BERTScore-F1~\cite{10443573} to measure the similarity between generated and ground-truth descriptions. 
Among them, BLEU and ROUGE mainly capture lexical overlap, METEOR offers more flexible word-level matching, and BERTScore-F1 evaluates semantic consistency based on contextual representations. 
At the odor-label level, we report facet coverage mean (FCM)~\cite{samuel2026coveragebench} and F1~\cite{xie2025multi} to assess whether the generated descriptions correctly cover each molecule's key odor labels.
Specifically, we use an LLM-based text embedding model to encode the generated description and each ground-truth odor label.
The cosine similarity between the description embedding and each label embedding is used to estimate the semantic relevance of that odor facet.
FCM measures the average coverage of ground-truth odor facets, while F1 reflects the balance between precision and recall in odor label prediction. 
These two groups of metrics jointly evaluate both the linguistic quality and odor-specific correctness of the generated descriptions.

\subsection{Quantitative Results}
As shown in Table~\ref{tab_quantitative_results}, our model achieves the best results across both the text-semantic level and the odor-label level, demonstrating its effectiveness in molecular odor description generation. 
Compared with general LLMs, our model shows clear advantages. 
Although strong general LLMs such as GPT-$5.5$ and Claude-Opus achieve relatively competitive results, they remain limited in capturing odor semantics from molecular inputs.
At the text-semantic level, our model improves over the best general LLM result on each metric by $171.5$\% in BLEU, $72.1$\% in METEOR, $25.6$\% in ROUGE, and $2.8$\% in BERTScore-F1.
At the odor-label level, it further raises the FCM and F1 scores by $3.1$\% and $7.1$\%, respectively. 
These results indicate that general LLMs can generate fluent descriptions, but they are less effective at accurately covering key odor facets.
Compared with the chemistry-oriented baseline ChemDFM, our model achieves more substantial improvements.
Although ChemDFM incorporates chemical knowledge, its performance on odor description generation remains limited, especially on text-level metrics. 
In contrast, our model improves over ChemDFM by $355.8$\% in METEOR, $178.0$\% in ROUGE, $22.0$\% in BERTScore-F1, $511.1$\% in FCM, and $445.1$\% in F1.
These results suggest that chemical knowledge alone is insufficient for this task.

Beyond the above prompting-based methods, SFT provides a stronger baseline because it is directly trained on the target molecular odor description generation task. 
Benefiting from supervised training on task-specific data, SFT can learn the mapping from molecular inputs to odor descriptions more effectively than general LLMs and chemistry-oriented models, which mainly rely on pretrained knowledge or prompting ability. 
As a result, SFT achieves the strongest baseline performance at both the text-semantic level and the odor-label level.
Nevertheless, our method still consistently outperforms SFT across all metrics. 
Specifically, compared with SFT, our model improves BLEU, METEOR, and ROUGE by $9.2$\%, $6.4$\%, and $10.3$\% at the text-semantic level, and further improves FCM and F1 scores by $10.2$\% and $13.7$\% at the odor-label level. 
These results indicate stronger molecule-to-odor semantic alignment, with descriptions closer to the ground-truth and broader coverage of key odor facets.

\subsection{Qualitative Results}

\begin{figure*}[!t]
    \centering
    \includegraphics[width=0.98\linewidth]{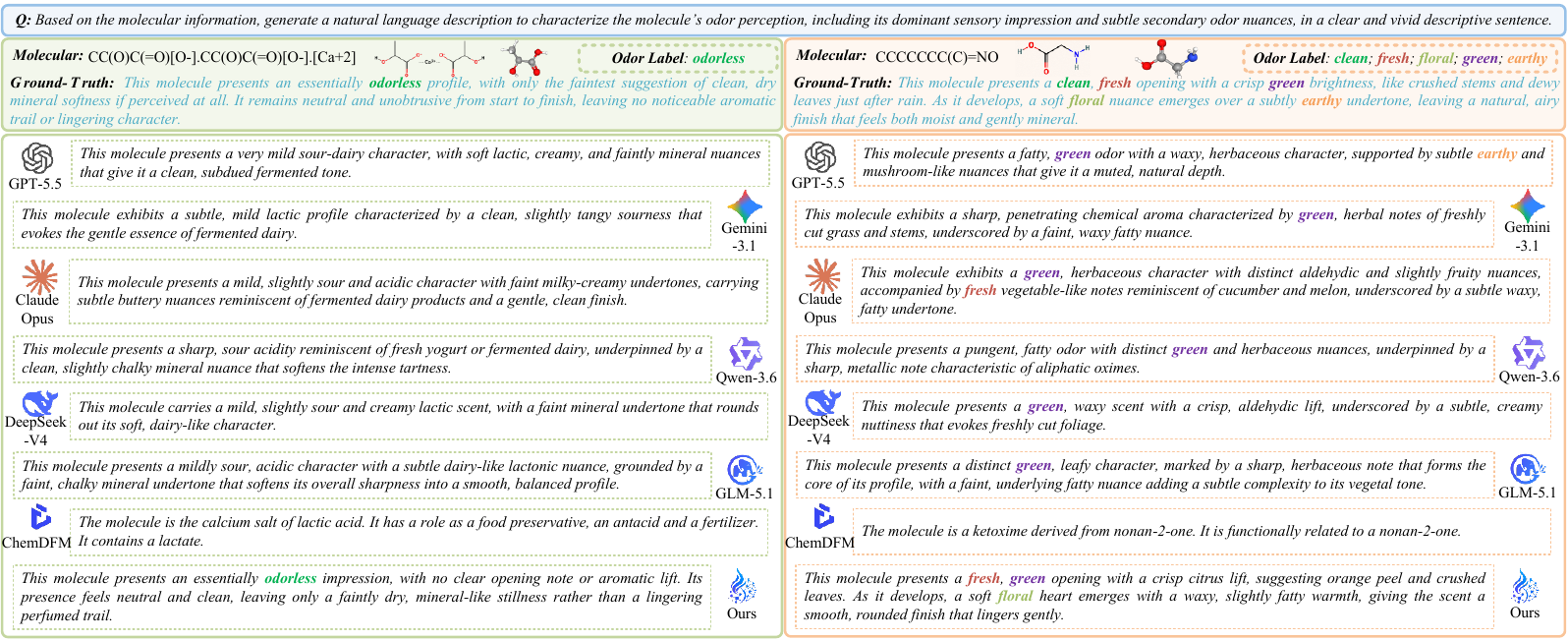}
    \vspace{-2pt}
    \caption{
    {Qualitative examples illustrating the task of molecular odor description generation.}
    }
    \label{fig_Qualitative}
    \vspace{-8pt}
\end{figure*}

Fig.~\ref{fig_Qualitative} presents two qualitative examples of molecular odor description generation, covering both single-label and multi-label cases. 
We compare our model with general LLMs and a chemistry-oriented model. 
Overall, existing models can generate fluent odor descriptions, but they often suffer from semantic drift or incomplete coverage of key odor facets. 
In contrast, our model produces descriptions that are more consistent with the ground-truth odor perception and provides richer sensory details, rather than merely listing odor labels.

In the single-label example, the target odor label is {odorless}, and the ground-truth description emphasizes an almost imperceptible profile with only faint clean, dry, and slightly mineral nuances. 
Several baselines fail to preserve this weak-odor property.
For example, some methods describe clear lactic, sour, creamy, buttery, or fruity impressions, which incorrectly amplify the odor intensity and introduce irrelevant sensory facets. 
The chemistry-oriented model also tends to focus on molecular functionality rather than perceptual odor semantics. 
By comparison, our model correctly captures the dominant odorless impression and further describes it as neutral, clean, and faintly mineral-like. 
These results show that our model can avoid over-generation when a molecule has only a subtle odor profile.

In the multi-label example, the target odor labels include {clean}, {fresh}, {floral}, {green}, and {earthy}. 
The ground-truth description presents a clean and fresh opening, green plant-like notes, a soft floral heart, and an earthy undertone.
Most baselines can capture part of the green or fresh character, but they either miss important secondary facets or introduce unrelated notes such as {chemical}, {metallic}, {aldehydic}, {fatty}, or {musky} impressions.
Our model generates a more complete and coherent description, describing a fresh and green opening with citrus, orange peel, and crushed leaves, followed by a soft floral heart and a smooth lingering finish.
Although the generated text does not explicitly reproduce all odor labels (\emph{e.g.}, clean and earthy), these attributes are reflected through semantically related expressions, such as crisp citrus lift, crushed leaves, and natural green freshness.
This demonstrates that our model can better generate natural and perceptually meaningful odor descriptions from molecular information.

\subsection{User Study}
\begin{figure}[!t]
    \centering
    \includegraphics[width=1\linewidth]{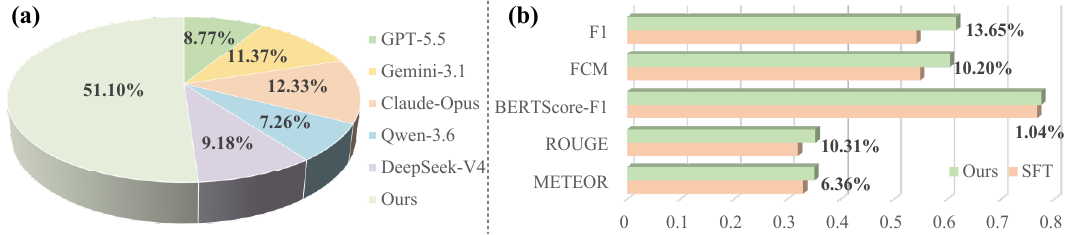}
    \caption{
    { (a) User study result. (b) Quantitative comparison with the SFT baseline.}
    }
    \label{fig_U}
\end{figure}

To further evaluate the quality of the generated odor descriptions from a human perspective, we conducted a user study by comparing our method with five strong LLM-based baselines, including GPT-$5.5$, Gemini-$3.1$, Claude-Opus, Qwen-$3.6$, and DeepSeek-V$4$. 
The questionnaire contained $10$ molecular cases.
For each case, participants were presented with the SMILES representation and a set of discrete odor labels.
They were asked to select the description that they considered to be (i) the most fluent, (ii) the most complete in covering the given odor labels, and (iii) the most factually accurate.
To ensure fairness, all candidate descriptions were anonymized, and their order was randomly shuffled for each question. 
We invited $73$ participants with backgrounds in chemistry or biology, and collected $730$ valid responses in total.
As shown in Fig.~\ref{fig_U} (a), our method received $51.10$\% of the votes.
This indicates that most participants regarded our descriptions as more fluent, more comprehensive in odor-label coverage, and more factually accurate.
This result further demonstrates the effectiveness of our method in the molecular odor description generation task.

\begin{table}[!t]
\centering
\footnotesize
\setlength{\tabcolsep}{8pt}
\caption{Quantitative comparison between odor semantic planner and existing baselines on the odor label prediction.}
\label{tab_micro_macro_results}
\vspace{-4pt}
\begin{tabular}{ccccc}
\hline
\hline
\multirow{2}{*}{Methods} & \multicolumn{2}{c}{Micro} & \multicolumn{2}{c}{Macro} \\
\cmidrule(lr){2-3} \cmidrule(lr){4-5}
 &AUROC$\uparrow$ & F1$\uparrow$ & AUROC$\uparrow$ & F1$\uparrow$ \\
\hline
\hline
POM~\cite{lee2023principal}      & 0.882 & 0.338 & 0.832 & 0.221 \\
Mol-PECO~\cite{zhang2024deep} & 0.840 & 0.274 & 0.780 & 0.148 \\
SM-LGB~\cite{suh2025comparative}   & 0.880 & 0.365 & 0.810 & 0.234 \\
Ours     &\textbf{0.923} &\textbf{0.475} &\textbf{0.865} &\textbf{0.248}\\
\hline
\hline
\end{tabular}
\vspace{-8pt}
\end{table}

\subsection{Ablation Study}

\subsubsection{\textbf{{Necessity of the Odor Semantic Planner}}}
\begin{figure}[!t]
    \centering
    \includegraphics[width=1\linewidth]{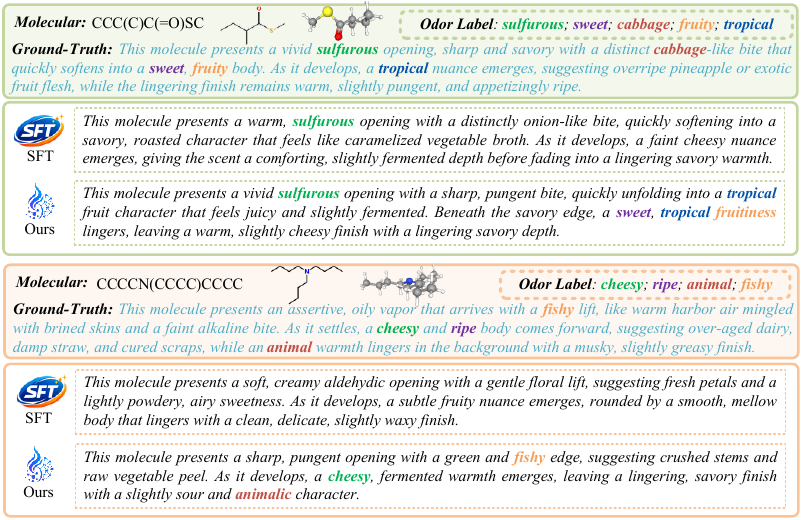}
    \caption{
    {Qualitative comparison with the SFT baseline.}
    }
    \label{fig_AQ}
    \vspace{-12pt}
\end{figure}

To verify the necessity of the odor semantic planner, we compare our model with an SFT baseline. 
This baseline is directly fine-tuned on the constructed MOD dataset, without any explicit odor semantic planning.
As shown in Fig.~\ref{fig_U} (b), our model consistently outperforms the SFT baseline on all evaluation metrics, improving F1, FCM, BERTScore-F1, ROUGE, and METEOR by $13.65$\%, $10.20$\%, $1.04$\%, $10.31$\%, and $6.36$\%, respectively.
These results show that direct supervised fine-tuning can generate fluent molecule-to-odor descriptions, but it is less effective in identifying and organizing key odor semantics from molecular information. 
In contrast, the odor semantic planner explicitly predicts informative odor cues before description generation.
These cues provide semantic guidance for the description generator, helping it generate descriptions that align more closely with the ground-truth odor perception. 
The qualitative examples in Fig.~\ref{fig_AQ} further support this finding. 
The SFT baseline can produce plausible descriptions, but often misses important odor attributes or emphasizes biased sensory impressions.
By comparison, our model better captures both dominant odor impressions and subtle secondary nuances, such as {sulfurous}, {fruity}, {cheesy}, and {animalic} characteristics. 
Therefore, the odor semantic planner is necessary for bridging molecular representations and natural odor descriptions, as it provides explicit semantic guidance and improves both odor-label consistency and descriptive richness.


\subsubsection{\textbf{{Effectiveness of the Odor Semantic Planner}}}
To evaluate the effectiveness of the proposed odor semantic planner, we compare it with three representative methods on the molecular odor label prediction task, including POM~\cite{lee2023principal}, Mol-PECO~\cite{zhang2024deep}, and SM-LGB~\cite{suh2025comparative}. 
As shown in Table~\ref{tab_micro_macro_results}, our odor semantic planner achieves the best performance across all evaluation metrics. 
These results show that the planner not only improves overall prediction accuracy, but also enhances the recognition of diverse odor categories. 
Compared with POM~\cite{lee2023principal}, which mainly learns a general perceptual odor map, our planner directly models label-level odor semantics and organizes molecular information into more discriminative odor concepts. 
Mol-PECO~\cite{zhang2024deep} focuses on molecular position encoding and electrostatic features, while our method further integrates multimodal molecular representations, which provide richer odor-related cues for semantic planning. 
SM-LGB~\cite{suh2025comparative} relies on molecular fingerprints and conventional machine learning models. 
In contrast, our planner captures higher-level semantic relations between molecular structures and odor labels, leading to stronger generalization.
These results demonstrate the effectiveness of the proposed odor semantic planner.

\subsubsection{\textbf{{Effect of Multimodal Molecular Inputs}}}
To evaluate the contribution of different molecular modalities, we conduct a modality ablation study on the odor semantic planner. 
Specifically, we train the planner with different molecular modalities as inputs and evaluate its performance on odor label prediction using micro-averaged and macro-averaged AUROC and F$1$ scores. 
As shown in Table~\ref{tab_modality_ablation}, each modality contributes to odor semantic prediction from a different perspective.
Among the single-modality settings, 3D molecular conformation achieves the best overall performance. 
In contrast, the 2D molecular graph alone shows relatively lower performance, suggesting that topological structure without sequential or spatial cues is insufficient to fully infer odor semantics.
When two modalities are combined, the performance generally improves over using a single modality, especially for the combination of 1D SMILES sequence and 2D molecular graph. 
This result indicates that chemical sequence information and structural topology provide complementary cues for recognizing odor-related molecular patterns.
The best performance is achieved when the 1D SMILES sequence, 2D molecular graph, and 3D molecular conformation are jointly used. 
These results demonstrate that the three molecular modalities are complementary, and their integration enables the odor semantic planner to learn a more comprehensive molecular representation.

\begin{table}[t]
\centering
\footnotesize
\caption{Ablation study on different molecular modalities.}
\vspace{-4pt}
\label{tab_modality_ablation}
\setlength{\tabcolsep}{4pt}
\begin{tabular}{ccccccc}
\hline
\hline
\multirow{3}{*}{SMILES} & \multirow{3}{*}{Graph} & \multirow{3}{*}{Conformation} 
& \multicolumn{2}{c}{Micro} & \multicolumn{2}{c}{Macro} \\
\cmidrule(lr){4-5} \cmidrule(lr){6-7}
 & & & AUROC$\uparrow$ & F1$\uparrow$ & AUROC$\uparrow$ & F1$\uparrow$ \\
\hline
\hline
\checkmark &  &  & 0.829 & 0.182 & 0.795 & 0.150 \\
 & \checkmark &  & 0.807 & 0.160 & 0.784 & 0.134 \\
 &  & \checkmark & 0.839 & 0.194 & 0.809 & 0.155 \\
\checkmark & \checkmark &  & 0.863 & 0.218 & 0.837 & 0.171 \\
\checkmark &  & \checkmark & 0.841 & 0.196 & 0.808 & 0.157 \\
 & \checkmark & \checkmark & 0.849 & 0.202 & 0.822 & 0.166 \\
\checkmark & \checkmark & \checkmark &\textbf{0.923} &\textbf{0.475} &\textbf{0.865} &\textbf{0.248} \\
\hline
\hline
\end{tabular}
\end{table}

\begin{table}[!t]
\centering
\footnotesize
\setlength{\tabcolsep}{8pt}
\caption{Ablation study on fusion strategies and loss $\mathcal{{L}}_{Cor}$. }
\vspace{-4pt}
\label{tab_fusion}
\begin{tabular}{ccccc}
\hline
\hline
\multirow{2}{*}{Methods} & \multicolumn{2}{c}{Micro} & \multicolumn{2}{c}{Macro} \\
\cmidrule(lr){2-3} \cmidrule(lr){4-5}
 &AUROC$\uparrow$ & F1$\uparrow$ & AUROC$\uparrow$ & F1$\uparrow$ \\
\hline
\hline
Concatenation      & 0.905 &0.419 &0.836 &0.220 \\
MoE &0.917 &0.425 &0.858 &0.246  \\
w/o $\mathcal{L}_{\mathrm{Cor}}$   &0.912 &0.434 &0.854 &0.231 \\
Ours     &\textbf{0.923} &\textbf{0.475} &\textbf{0.865} &\textbf{0.248}\\
\hline
\hline
\end{tabular}
\vspace{-8pt}
\end{table}

\subsubsection{\textbf{{Ablation on Sparse MoE and Loss $\mathcal{L}_{\mathrm{Cor}}$}}}
To evaluate the fusion strategy and the loss $\mathcal{L}_{\mathrm{Cor}}$, we conduct an ablation study in Table~\ref{tab_fusion}. 
We first test simple concatenation of multimodal features. This variant yields the lowest performance, indicating that directly merging features fails to capture complementary odor cues.
We then replace concatenation with a plain MoE. The performance improves, showing that expert-based fusion is more effective than fixed concatenation, since it models both modality-specific and cross-modal patterns. 
However, this variant activates all experts, which makes the fusion less targeted.
In contrast, our sparse MoE selects the two most relevant experts for each molecule, focusing on informative modality interactions while suppressing fusion noise, thereby achieving the best overall performance.
Finally, we remove loss $\mathcal{L}_{\mathrm{Cor}}$ to examine its contribution. 
The performance drops consistently across all metrics. 
This shows that $\mathcal{L}_{\mathrm{Cor}}$ helps the planner learn label co-occurrence relations among odor categories, leading to more accurate and consistent odor semantic prediction.

\subsubsection{\textbf{{Necessity of the Semantic Adapter}}}
\begin{figure}[!t]
    \centering
    \includegraphics[width=1\linewidth]{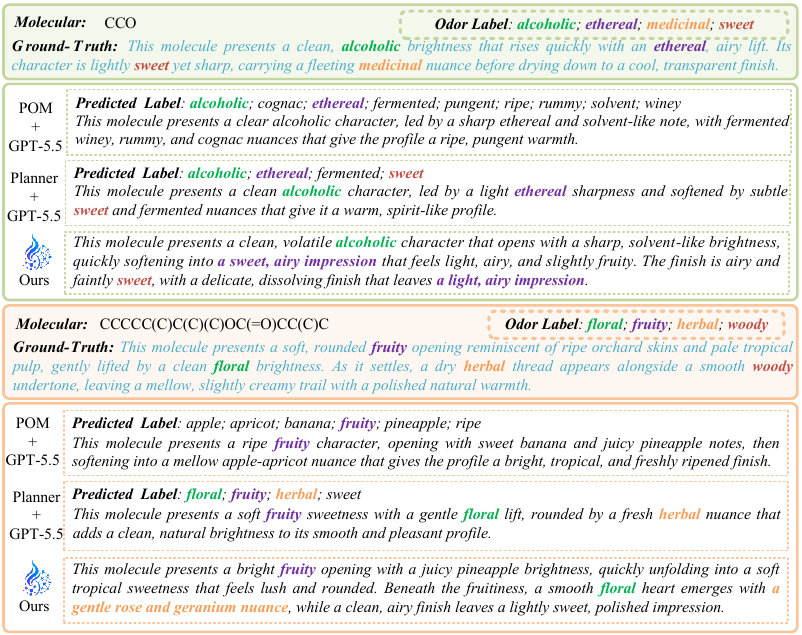}
    \vspace{-13pt}
    \caption{
    {Qualitative comparison of our model and adapter-free two-stage baselines.}
    }
    \label{fig_AA}
    \vspace{-10pt}
\end{figure}

To verify the necessity of the semantic adapter, we compare our model with two adapter-free two-stage baselines. 
In these baselines, POM~\cite{lee2023principal} or our odor semantic planner first predicts discrete odor labels, and GPT-5.5 then generates the full odor description using only these labels. 
As shown in Fig.~\ref{fig_AA}, this simple strategy can produce fluent descriptions, but it heavily relies on the quality and completeness of the predicted labels. 
When the labels are noisy, missing, or too coarse, GPT-5.5 may generate natural descriptions that are less consistent with the ground-truth odor perception.
For example, POM~\cite{lee2023principal} may introduce less relevant labels, such as {cognac}, {pungent}, {ripe}, {rummy}, {solvent}, and {winey}, leading to over-expanded descriptions.
It may also miss important cues, such as {floral}, {herbal}, and {woody} notes, resulting in incomplete descriptions. 
The planner-based baseline predicts cleaner labels, but it still treats them as isolated words. 
As a result, its descriptions are often limited to general impressions, such as {alcoholic} or {ethereal} profiles, and may fail to express richer perceptual details. 
In contrast, our model generates more coherent and complete descriptions by capturing key odor characteristics, including volatile alcoholic notes, solvent-like brightness, fruity openings, pineapple brightness, tropical sweetness, floral nuances, and a sweet polished finish.
Notably, our model does not simply copy the predicted labels.
It can express the same odor semantics with more natural phrases.
For example, an {ethereal} note can be described as an airy impression or a light, volatile brightness, while a {floral} note can be expressed as a rose and geranium nuance. 
This shows that the semantic adapter does not merely pass hard labels to the language model. 
It converts odor semantics into continuous soft prompts, which preserve richer semantic relations and enable more flexible language generation. 
In this way, the semantic adapter better aligns molecular information, odor semantics, and natural odor descriptions, while reducing the error propagation caused by hard-label-based two-stage generation.

\section{Conclusion}
In this paper, we proposed ScentGen, a hierarchical multimodal olfactory semantic modeling framework for molecular odor description generation. 
We also constructed a customized molecular odor description dataset to support this new research direction.
Extensive experiments demonstrate that ScentGen can generate coherent and expressive odor descriptions, providing a more flexible solution for molecular odor understanding beyond fixed odor label prediction.
We hope this work will encourage further research on language-based molecular odor modeling and multimodal olfactory intelligence.

\bibliographystyle{IEEEtran}
\bibliography{ref}

\end{document}